\documentclass[proof]{pasj00}
\usepackage{txfonts}
\usepackage{natbib}
\bibpunct{(}{)}{;}{a}{}{,}
\usepackage[dvips]{graphicx}

\begin{document}

\title{Probing the Accretion Scheme of the Dipping X-ray Binary 4U 1915-05 with Suzaku}

\author{Zhongli \textsc{Zhang}\altaffilmark{1}, Kazuo \textsc{Makishima}\altaffilmark{1,2,3}, Soki \textsc{Sakurai}\altaffilmark{1}, Makoto \textsc{Sasano}\altaffilmark{1}, and Kou \textsc{Ono}\altaffilmark{1}} 
\altaffiltext{1}{Department of Physics, School of Science, The University of Tokyo, 7-3-1, Hongo, Bunkyo-ku, Tokyo 113-0033, Japan}
\altaffiltext{2}{MAXI Team, Institute of Physical and Chemical Research (RIKEN), Wako, Saitama 351-0198, Japan}
\altaffiltext{3}{Research Center for the Early Universe, The University of Tokyo, 7-3-1, Hongo, Bunkyo-ku, Tokyo 113-0033, Japan}
\email{zzhang@juno.phys.s.u-tokyo.ac.jp}


\maketitle


\begin{abstract}

The dipping low-mass X-ray binary 4U 1915-05 was observed by Suzaku on 2007 November 8 for a net exposure of 39 ksec. It was detected by the XIS with a 0.8-10 keV signal rate of $9.84\pm0.01$ cts s$^{-1}$ per camera, and HXD-PIN with a 12-45 keV signal rate of $0.29\pm0.01$ cts s$^{-1}$. After removing the periodic dips and an X-ray burst, the 0.8 -- 45 keV continuum was successfully described by an optically thick disk emission with an inner-disk temperature $\sim0.7$ keV and a neutron-star blackbody emission with a temperature $\sim1.3$ keV, on condition that the blackbody component, or possibly the disk emission too, is significantly Comptonized. This successful modeling is consistent with 4U 1915-05 being in a high-soft state in this observation, and implies that its broadband spectrum can be interpreted in the same scheme as for many non-dipping Low-mass X-ray binaries in the soft state. Its bolometric luminosity ($\sim$ 0.02 times the Eddington limit) is relatively low for the soft state, but within a tolerance, if considering the distance and inclination uncertainties. As a high-inclination binary, this source exhibited stronger Comptonization effect, with a larger Comptonizing $y$-parameter, compared to low and medium inclination binaries. This suggests that the Comptonizing coronae of these objects in the soft state is in an oblate (rather than spherical) shape, extending along the accretion disk plane, because the $y$-parameter would not depend on the inclination if the corona were spherical.

\end{abstract}

\section{INTRODUCTION}
\label{sec:introduction}

A neutron-star low-mass X-ray binary (NS-LMXB, or hereafter simply LMXB) consists of a mass accreting neutron star (NS) and a mass-donating low-mass star which fills the binary Roche lobe. It is well known that these objects show bimodality in X-ray spectral states, namely soft and hard states \citep[e.g.,][]{White85,Mitsuda84,Mitsuda89}. In the soft state when the mass accretion rate is high, mass from the companion spirals onto the NS through a standard accretion disk \citep{Shakura73} which is optically thick and geometrically thin. The spectrum is soft, which we can describe by the canonical model of \citet{Mitsuda84}: a disk blackbody with a temperature of $\sim$1 keV, plus a blackbody from the NS surface with a temperature of $\sim$ 2 keV. The latter component is often found to be weakly Comptonized \citep[e.g.,][]{Mitsuda89,Barret01,Iaria05}. Over the past three decades, this model has been reinforced by a number of studies including, e.g., \citet{Makishima89}, \citet{Farinelli03}, \citet{Takahashi11}, \citet{Sakurai12} and \citet{Muck13}. When the accretion rate is low and the source is in the hard state, the inner accretion disk transites into an optically thin and geometrically thick flow around the NS, which is further characterized by a large radial velocity component, a nearly free-fall ion temperature, and possibly a significant difference between the ion and electron temperatures. The hot electrons therein strongly Comptonize the NS blackbody emission and harden the spectrum to $\gtrsim$ 100 keV \citep{Mitsuda89,Sakurai12}. 

In contrast to these ``normal" objects, a special kind of LMXBs with high inclination angles, namely dipping LMXBs (also called dippers) showing periodic dips in their X-ray intensity, seldom have their spectral states investigated. For a long time, their dips were thought to be caused by progressive covering of a largely extended scattering region called accretion disk corona \citep[ADC,][]{White82b} by the companion star \citep[e.g.,][]{Church98,Morley99}. However, new studies revealed that the dips can be explained simply by obscuration of the central compact X-ray source by an ionized and extended absorber located at some outer parts of the accretion disk \citep{Trigo06}, and the ADC is no longer required. Based on the new scenario, it is reasonable to consider dipping LMXBs essentially the same objects as non-dipping LMXBs, except for their higher inclinations. Hence it is worthwhile to examine whether we can understand dippers in the same spectral classification scheme as for non-dipping LMXBs, possibly with some modifications.

So far, broad-band spectra of dipping LMXBs were studied extensively, e.g., with BeppoSAX. Examples include relatively luminous dippers 4U 1915-05 \citep{Church98}, XB 1254-690 \citep{Iaria01} and XB 1658-298 \citep{Oosterbroek01}, as well as dimmer ones such as XB 1323-619 \citep{Church99} and EXO 0748-676 \citep{Sidoli05}. The reported spectra are generally rather hard, often extending to $\sim$ 100 keV, which are reminiscent of the hard state of non-dipping LMXBs. However, these previous works did not clearly specify whether the objects were in the hard state or not. Since dippers, like the other LMXBs, must be distributed over the two states, we presume that some of them actually reside in the soft state, and their broad-band spectra appear harder because, e.g., their higher inclination leads to a stronger Comptonization. To examine this possibility, we need to analyze broad-band non-dipping spectra of some appropriate dippers in a unified way which is consistent with the study of non-dipping objects. The available publications are clearly inadequate for this purpose, because different authors employed different modelings.

For the above purpose, we need to use an X-ray instrument with a high spectral resolution and broad-band coverage. The most suited one is the Japanese satellite Suzaku \citep{Mitsuda07}, which has a soft X-ray energy resolution of $\sim100$ eV realized by the X-ray Imaging Spectrometer \citep[XIS;][]{Koyama07}, and an energy coverage up to $>$ 100 keV achieved by the hard X-ray detector \citep[HXD;][]{Takahashi07}. So far, five dipping LMXBs have been observed in total with Suzaku, and among them, we selected the most luminous one 4U 1915-05 (see section \ref{sec:observation} for details). This source is the first dipper discovered in 1982 \citep{White82}, showing stable periodic dips with a period of $P_{\rm s}\sim 50$ mins \citep{White82,Homer01}, and frequent type I X-ray bursts \citep{Swank84}. It is also known as an ultracompact X-ray binary (UCXB) which has a helium rich donor star \citep{Nelemans06}. Its 0.6-10 keV persistent luminosity, was $\sim 4.4 \times 10^{36}$ erg s$^{-1}$ \citep{Balman09} at an assumed distance of 9.3 kpc (see below), reached the range typical of LMXBs in the soft state. The broadband spectrum of this source was studied by \citet{Church98} using BeppoSAX. By fitting a 0.5-200 keV  spectrum with a cut-off power-law model, these authors reported a rather high cut-off energy of $\sim$ 80 keV, which is suggestive of strong Comptonization. However, further exploration with other X-ray satellites (e.g., ROSAT, Chandra, XMM-Newton) did not give much improved knowledge of this issue due to their limited energy coverage \citep{Morley99,Boirin04,Iaria06}. Furthermore, few of these studies attempted spectral classification of this object in comparison with non-dipping LMXBs. 
  
In the present paper, we utilized an archival Suzaku data set of 4U 1915-05 acquired on 2006 November 8. We focus on its non-dip broadband spectral properties, while the X-ray dips are out of the scope of our study. Throughout the paper, the source distance is assumed to be 9.3 kpc, as determined by \citet{Yoshida93} with an uncertainty of $\sim 15$\%, using the X-ray bursts as a standard candle. Using the same method, other studies gave similar values of 8.4-10.8 kpc \citep{Smale88} and 7-9 kpc \citep{Galloway08}; thus, the distance uncertainty is typically 20\%. Since the source shows X-ray dips, but no X-ray eclipse, its inclination angle is estimated to be $\theta\sim70^{\circ}$ \citep{Vanpara88}, with a $\pm10^{\circ}$ uncertainty \citep{Frank87}.

\section{OBSERVATION AND DATA PROCESSING}
\label{sec:observation}

Table \ref{tab:fourdippers} summarizes the Suzaku observations of four dipping LMXBs among the five sources in the archive, excluding 4U 1822-37 which has been found to have strong magnetic field \citep[$\sim$ 12 G;][]{Sasano14}. We estimated their 1-50 keV luminosities, by fitting their background subtracted spectra with an absorbed cut-off power-law model, and then removing the absorption. The bolometric corrections, which depend on the cutoff energy, are typically smaller than 1.3. Thus, among the four dippers observed with Suzaku, 4U 1915-05 is the most luminous, and hence is suited to our purpose of identifying a dipping LMXB in the soft state. 

\begin{table}
\begin{center}
\caption{Suzaku observations of four dipping LMXBs.}
\label{tab:fourdippers}
\begin{tabular}{lcccc}
\hline
\hline

Dippers         &  $d^{\ast}$   &  Obs.ID     &  Exp. & $L_{\rm X}^{\dag}$  \\
                & (kpc)  &             &   (ks)       & ($10^{36}$ erg s$^{-1}$)  \\ 
\hline

4U 1915-05      &  9.3   &  401095010  &  39 &   5.3    \\ 
XB 1323-619     &  10    &  401002010  &  50 &   2.9    \\ 
EXO 0748-676    &  7.1   &  402092010  &  55 &   2.9    \\
XTE J1710-281   &  16    &  404068010  &  50 &   3.1    \\ 
\hline
\end{tabular}
\end{center}  
$^{\ast}$ The source distances taken from \citet{Yoshida93}, \citet{Church99}, \citet{Galloway08b}, and \citet{Younes09}, in the order of the sources.\\ 
$^{\dag}$ Unabsorbed 1-50 keV persistent luminosity of each source. 
\end{table}

The present Suzaku observation of 4U 1915-05 was conducted in 2006, starting at 05:49:53 on November 8 (54047 MJD) and ending at 02:41:15 on November 9, for a total elapsed time of 73.9 ksec and a net exposure time of 39.1 ksec. The observation was carried out by all four XIS detectors (numbered from 0 to 3), together with the HXD-PIN and HXD-GSO detectors. However, during this observation around 01:04:15 on November 9, XIS 2 stopped working. Thus, among the four XIS detectors, we utilized XIS 0 and XIS 3 which employ front-illuminated (FI) CCDs and have similar responses. In the hard X-ray band, we analyzed only HXD-PIN data, because the source was not detected with HXD-GSO. Both the XIS and HXD-PIN data were analyzed by HEAsoft version 6.13, which includes \textit{Ftools}, \textit{Xanadu} and \textit{Xselect} packages. The Suzaku calibration (version 20070731 for XIS and version 20070710 for HXD-PIN) was used by the time of our data processing.

\subsection{XIS and HXD Data Processing}
\label{sec:analysis}

We co-added the data from the two FI detectors, XIS 0 and 3, which were both operated in normal mode, using the full window option with a time resolution of 8 s. Events with GRADE 0,2,3,4 and 6 were selected, including both $3\times3$ and $5\times5$ pixel options. Since the average count rate of each XIS camera, $>$ 10 cts s$^{-1}$, is high enough to cause pile-up effects at the source center, we excluded the central $0^{\prime}$.5 in order to reduce the pile-up effect to $<$ 3\% \citep{Yamada12}. Then, the source events for the spectral analysis were obtained from an annulus with the inner and outer radii of $0^{\prime}$.5 and $4^{\prime}$, respectively, while the background events were obtained from an outer annulus from $4^{\prime}$.5 to $6^{\prime}$.5. The background-subtracted average XIS 0 and XIS 3 source count rate is $9.84\pm0.01$ cts s$^{-1}$ in 0.8-10 keV.  

The raw HXD-PIN count rate in 12-45 keV was $0.83\pm0.01$ cts s$^{-1}$, of which $\sim~60\%$ was Non X-ray Background (NXB) according to the NXB model provided by the HXD team, and $\sim~3\%$ was the Cosmic X-ray Background (CXB). After subtracting these two background components, we are left with a 12-45 keV signal count rate of $0.29\pm0.01$ cts s$^{-1}$. The source was undetectable above 45 keV.

\subsection{Lightcurves}
\label{sec:lightcurve}

\begin{figure}
\begin{center}
\resizebox{0.75\hsize}{!}{\includegraphics[angle=0]{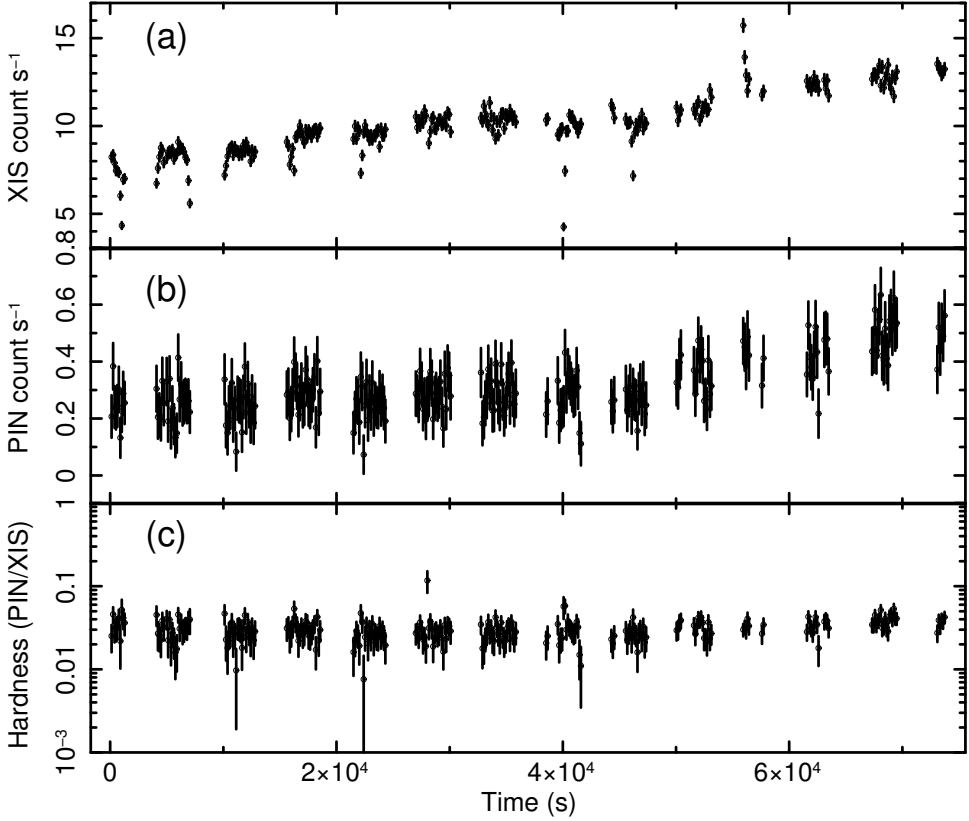}}
\caption{Suzaku lightcurves of 4U 1915-05 with 128 s bins. (a) The 0.8-10 keV average lightcurve of XIS 0 and XIS 3, with background subtracted. In addition to the dips, a type I X-ray burst is seen at $5.6\times10^{4}$ s. (b) NXB-subtracted HXD-PIN lightcurve in 12-45 keV. The CXB contribution by $\sim$ 0.025 cts s$^{-1}$ is included. (c) HXD-PIN vs. XIS hardness ratio.}
\label{fig:lightcurve}
\end{center}
\end{figure}

In the analysis of lightcurves (and of spectra in Section \ref{sec:spectra}), we filtered the exposure times of the XIS and HXD-PIN data with a common good time interval (GTI) file, and obtained a total common exposure of 34.6 ksec, which is 88\% of the original exposure of the XIS. From the HXD-PIN data, only the NXB was subtracted, because the CXB is not variable and does not affect the timing analysis. Figure \ref{fig:lightcurve} shows the derived XIS and HXD-PIN lightcurves, and the hardness ratio between them. The observation spans over 24 orbital periods of the binary, within which we detected around 10 dips (with the rest falling in the data gaps) and one type I X-ray burst; however, their detailed analysis is out of the scope of this paper. Except for these dip and burst events, the XIS count rate increased slowly and steadily, by up to a factor of 2. A similar source brightening is also observed in the HXD-PIN lightcurve. Within 1$\sigma$ errors, the PIN/XIS hardness ratio remained unchanged.

\section{THE BROAD-BAND NON-DIP SPECTRAL FITTING}
\label{sec:spectra}

We investigated the broad-band non-dip spectrum by simultaneously fitting the XIS and HXD-PIN data. Since the PIN/XIS hardness ratio remained approximately constant (figure \ref{fig:lightcurve}), the whole observation period was utilized. After excluding the dips and the burst from the data, the remaining exposure time became 25.1 ksec. We generated the response and arf files of the XIS using \textit{xisrmfgen} and \textit{xissimarfgen}, respectively. The response file of HXD-PIN, released officially from the HXD team,  was utilized with the option of ``XIS nominal". We used XSPEC (version 12.8.0) and the models therein for the spectrum fitting. The CXB contribution for HXD-PIN was included in a model as a fixed component as
\begin{equation}
\textmd{CXB}(E) = 9.41\times 10^{-3}  \left(\frac{E}{1\textmd{keV}}\right)^{\hspace{-0.3em}-1.29}\hspace{-1.4em}\exp\left(-\frac{E}{40\textmd{keV}}\right)   \label{eq:cxb}
\end{equation}
\citep{Boldt87}, where the unit is photons cm$^{-2}$ s$^{-1}$ keV$^{-1}$ FOV$^{-1}$. The cross normalization of HXD-PIN relative to the XIS was fixed at 1.158 \citep{Kokubun07}. Energy ranges of 0.8-10 keV and 12-45 keV were chosen to analyze the XIS and HXD-PIN data, respectively. To avoid the calibration uncertainties around the instrumental silicon K-edge and the gold M-edge, we excluded two XIS energy ranges, 1.7-1.9 keV and 2.2-2.4 keV, respectively. The calibration uncertainty of Al K-edge was fixed by adding an \textit{edge} model with energy fixed to 1.56 keV.

\begin{figure}
  \begin{center}
    \includegraphics[angle=0, width=0.48\textwidth]{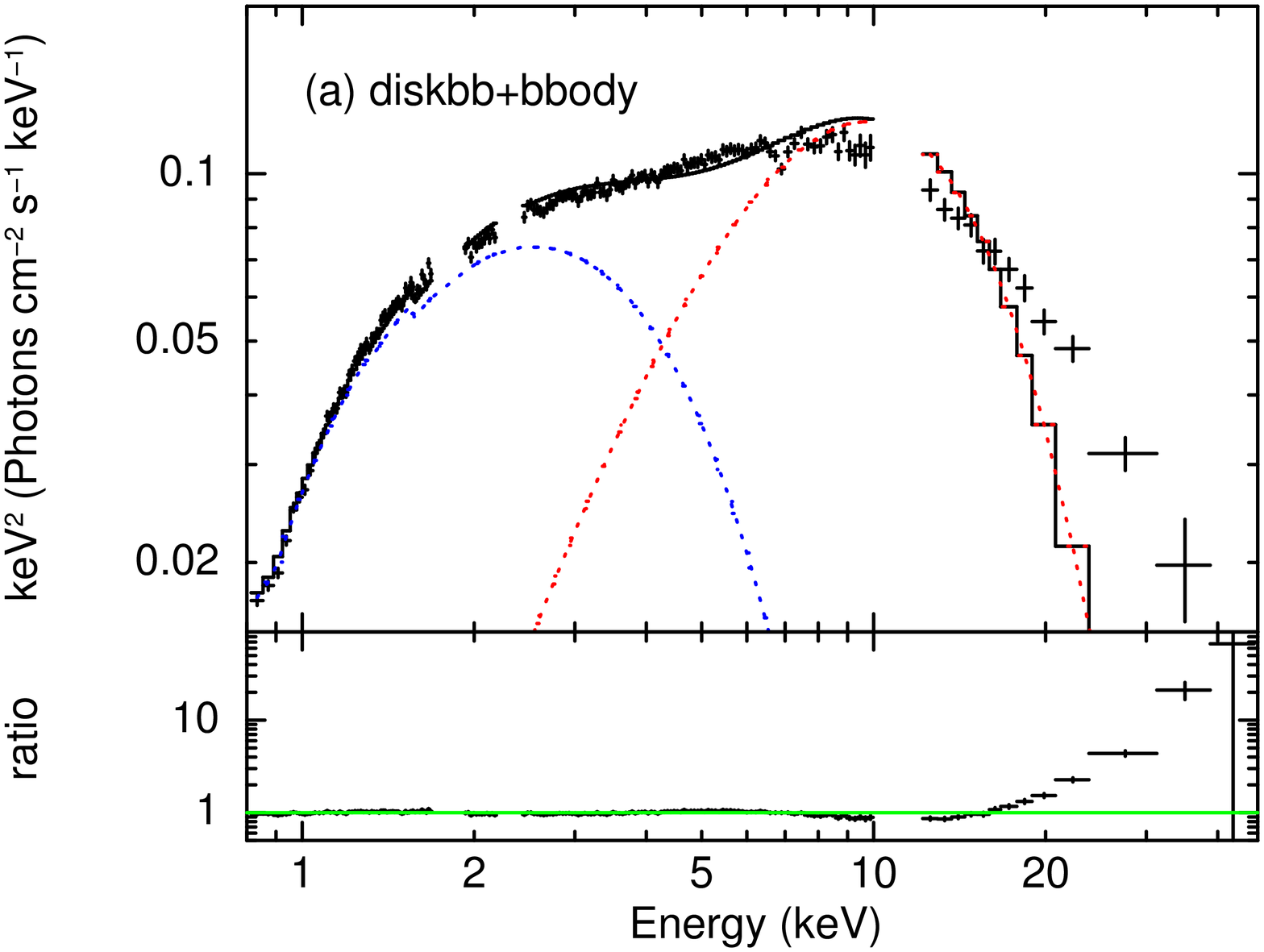}
    \includegraphics[angle=0, width=0.48\textwidth]{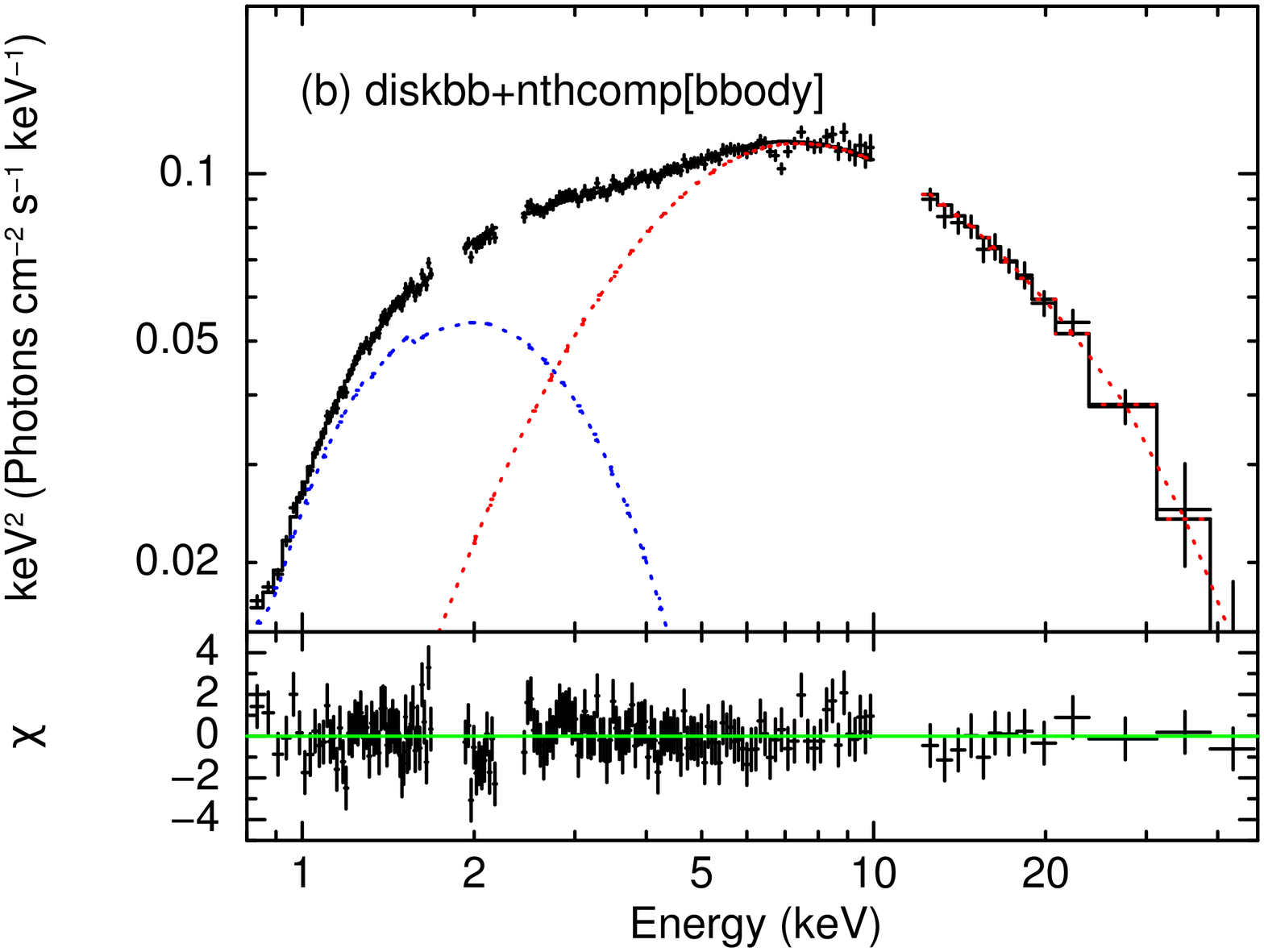}
    \includegraphics[angle=0, width=0.48\textwidth]{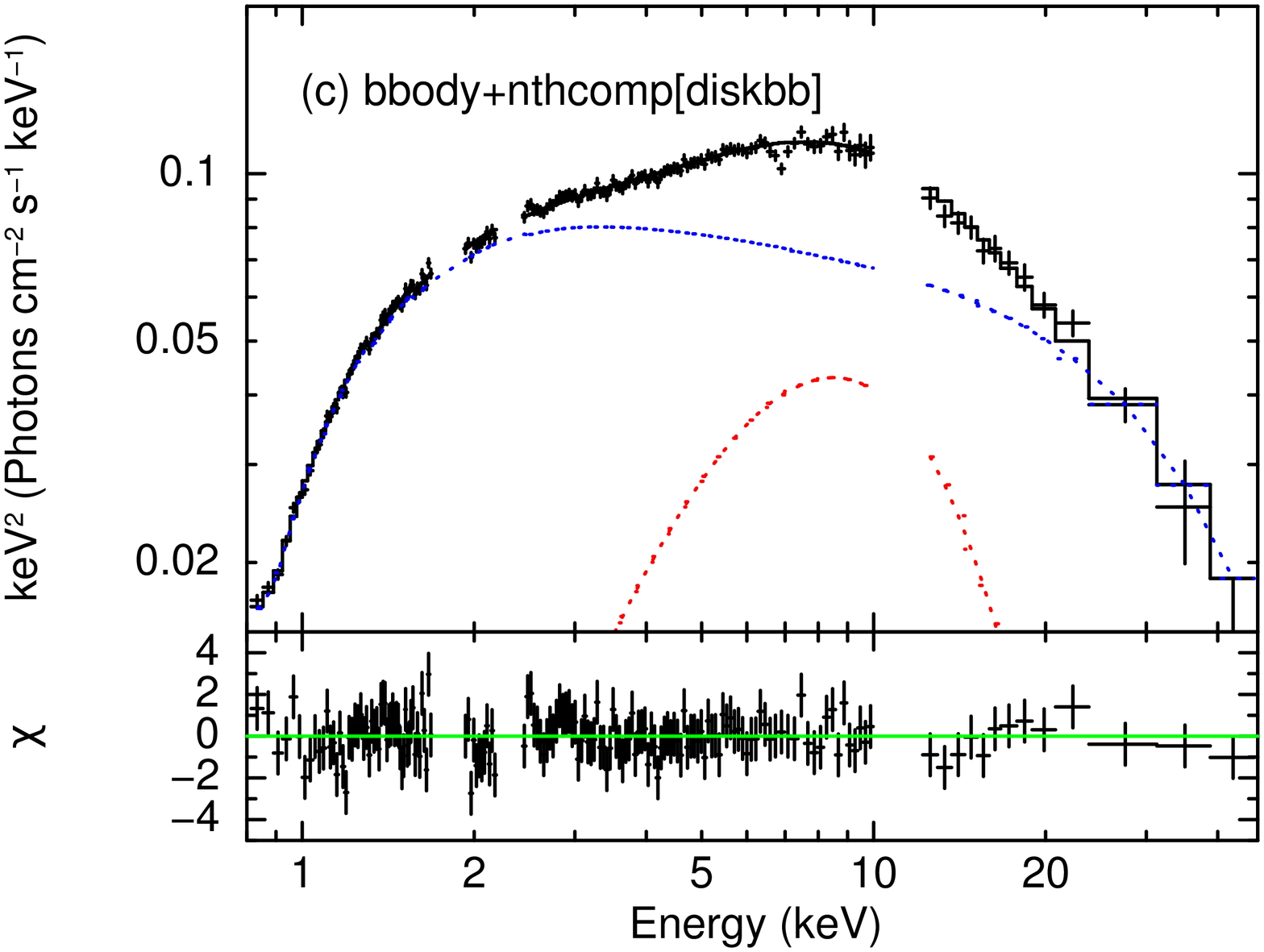}
    \includegraphics[angle=0, width=0.48\textwidth]{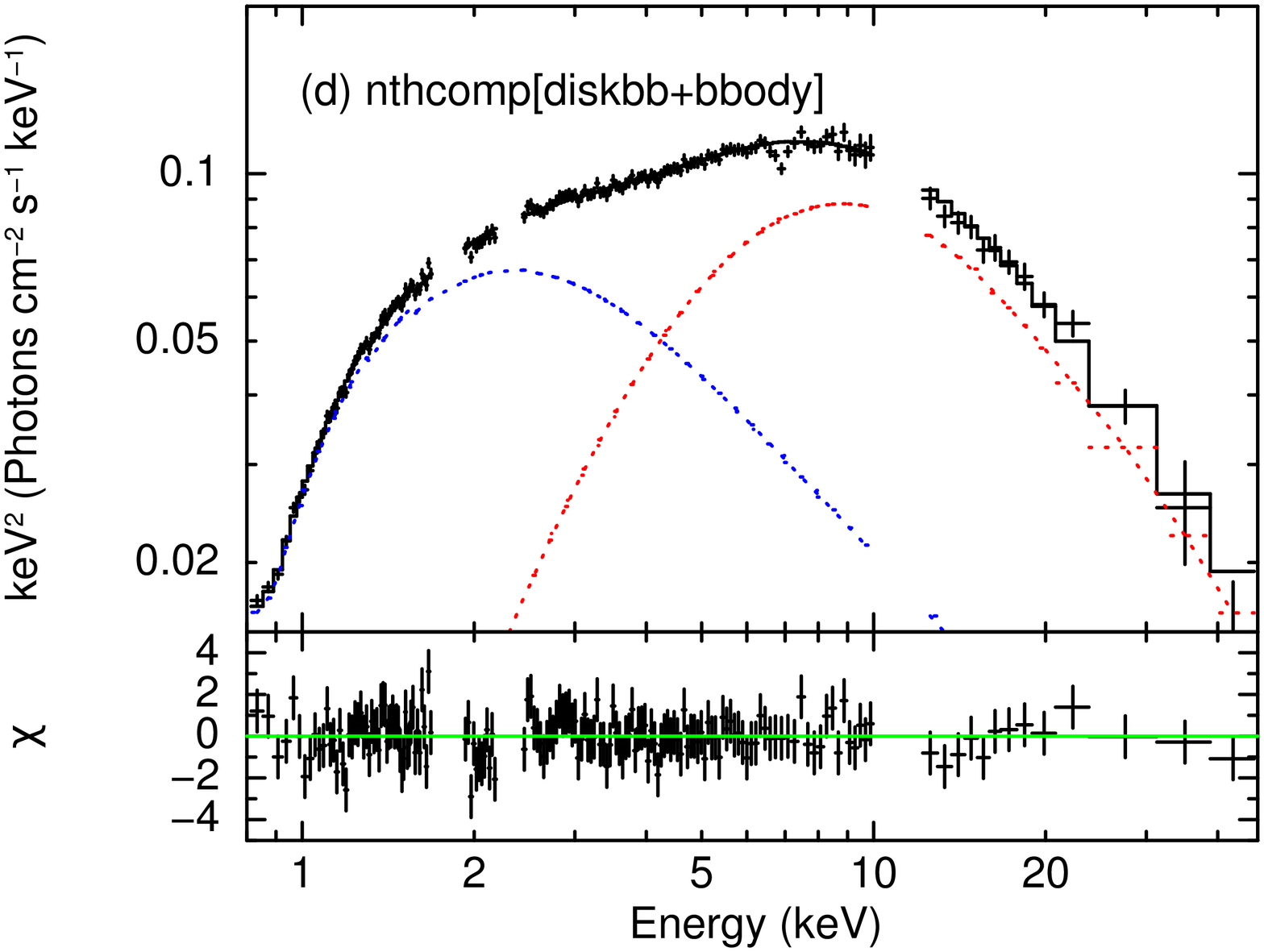}
    \end{center}
  \caption{The $\upsilon F \upsilon$ plot of the time-averaged non-dip Suzaku spectrum of 4U 1915-05 and its model fitting results. For the best-fit model, only its continuum components are shown. When deconvolving the spectra, the absorption-line factors were removed from the model, in order to avoid the ``obliging" effects on them. (a) Fit with a {\tt diskbb} (blue) plus {\tt bbody} (red) model and the ratio of the data to the model. (b) Fit with a {\tt diskbb} (blue) plus {\tt nthcomp[bbody]} (red) model and its residuals. (c) Fit with a {\tt bbody+nthcomp[diskbb]} model and its residuals. (d) Fit with a {\tt nthcomp[diskbb+bbody]} model and its residuals.}
\label{fig:fourspectra}   
\end{figure}

Figure \ref{fig:fourspectra} shows the derived spectrum in the $\nu F\nu$ form, together with the model fitting results. Thus, the spectrum rises quickly from $\sim$ 1 to $\sim$ 2 keV, keeps a relatively constant slope (with a photon index of $\sim$ 1.7) between $\sim$ 2 and $\sim$ 10 keV, and decreases steeply above $\sim$ 10 keV. In addition, some absorption lines are seen in the energy range of 6.5-7.5 keV. These may be identified with the Fe {\small XXV} K$\alpha$ and Fe {\small XXVI} K$\alpha$ absorption lines, previously detected at 6.65 and 6.95 keV respectively \citep[e.g.,][]{Boirin04}. Although these absorption features are expected to provide valuable information, we do not go into details of this issue, since our purpose here is to study the continuum. 

\subsection{Fit with a {\tt diskbb+bbody} Model} 
\label{sec:model1}

As shown in figure \ref{fig:fourspectra}, the $\nu F \nu$ spectrum is much more convex than those of typical LMXBs in the hard state, which are approximately flat in the $\nu F \nu$ form, and is more similar to those in the soft state \citep[e.g.,][]{Sakurai12}. Thus, we attempted to fit the data with the canonical model of \citet{Mitsuda84}; as already described in section \ref{sec:introduction}, it consists of a multi-color disk blackbody component ({\tt diskbb}) and a blackbody component ({\tt bbody}) from the NS surface. In the fitting we included the instrumental Al K-edge, and the celestial Fe absorption lines by multiplying two absorption gaussian ({\tt gabs}) components on the continuum models, but only show the {\tt diskbb+bbody} continuum in figure \ref{fig:fourspectra}(a). This model reproduced the spectra well up to $\sim$ 15 keV, with the blackbody temperature of $kT_{\rm bb}\sim$ 2.5 keV, its radius of $R_{\rm bb}\sim$ 0.8 km, the inner disk temperature of $kT_{\rm in}\sim$ 1.0 keV and its corresponding radius of $R_{\rm in}\sim 5.9$ km ($\theta =70^{\circ}$). However, the fit was not acceptable with $\chi^2(\nu)=921.1(159)$, with large positive residuals above 15 keV. This suggests that the canonical model for the LMXB spectra is significantly modified with inverse-Compton effects.

\subsection{Fit with a Single-Seed Comptonization Model} 
\label{sec:model2}

\begin{figure*}
  \begin{center}
    \includegraphics[angle=0, width=1.0\textwidth]{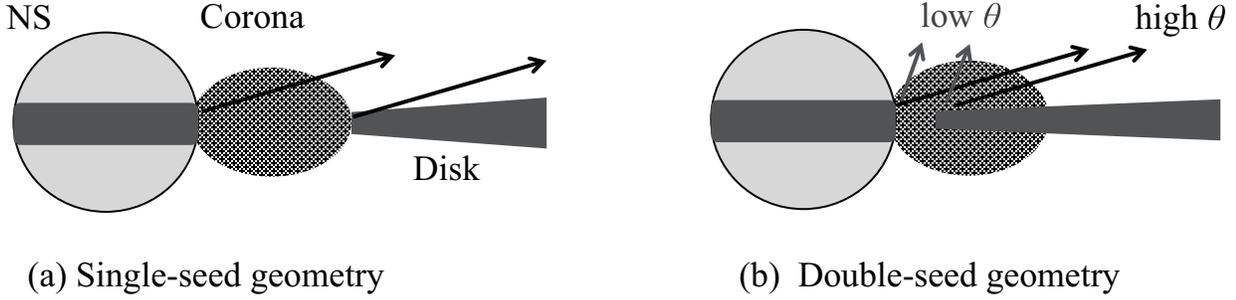} 
  \end{center}
  \caption{Conceptual illustrations of the accretion configuration of NS-LMXBs in the soft state. The X-ray emission arises from the NS surface and the inner accretion disk, and the Comptonizing corona is located between them in two possible geometries. (a) Seed photons are supplied only from the NS surface, while the disk is truncated at the outer boundary of the corona. (b) The disk intrudes into the corona, so that the two components are both Comptonized by a common corona with (possibly) different optical depths. In both configurations, high inclination sources are considered more strongly Comptonized due to the oblate coronal shape.}
\label{fig:geometry}    
\end{figure*}

To better describe the spectrum at $>$ 15 keV in terms of Comptonization, we may assume that only the blackbody photons from the NS surface are Comptonized, or that the disk photons are also Compton scattered. A conceptual illustration of these two configurations are shown in figure \ref{fig:geometry}. As the simpler of them, we applied Comptonization only to the NS blackbody component as suggested in, e.g.,  \citet{Mitsuda89} and \citet{Sakurai12}. For the Comptonization code we employed {\tt nthcomp} \citep{Zdziarski96,Zycki99}, which is flexible in choosing the seed photon source. Hence the model for the fitting is {\tt diskbb+nthcomp[bbody]}, where the square brackets specify the seed photon source. Like in subsection \ref{sec:model1}, we included the Al K-edge and Fe absorption lines in the model. As shown in table \ref{tab:totalspectra} and figure \ref{fig:fourspectra}(b), the fit has become successful with $\chi^2(\nu)=155.5(157)$, because the positive residuals in figure \ref{fig:fourspectra}(a) disappeared. As detailed in table \ref{tab:totalspectra}, the {\tt nthcomp} component has been characterized by a coronal electron temperature of $kT_{\rm e} \sim 9$ keV, and a spectral slope $\Gamma \sim 2.5$. Using the equation
\begin{eqnarray}
\Gamma &=& -\frac{1}{2} + \sqrt{\frac{9}{4}+\frac{1}{\frac{kT_\textmd{\scriptsize{e}}}{m_\textmd{\scriptsize{e}}c^2 }\tau(1+\frac{\tau}{3})}}
\end{eqnarray}
\citep{Lightman87}, the optical depth of the Comptonizing corona is estimated as $\tau \sim 4$.   

Physical conditions implied by the best-fit parameters in table \ref{tab:totalspectra}, including $kT_{\rm in} < kT_{\rm bb}$ and $R_{\rm bb} <$ 10 km, are all reasonable for a standard disk-accretion geometry and a single-seed Comptonization assumption. However, the condition of $R_{\rm in} \sim$ 10 km, namely the disk continuation down to the NS surface, may contradict the single-seed assumption, because photons from the inner disk region would then be Comptonized as well (to be discussed in section \ref{sec:discuss2}). This possibility of double-seed photon Comptonization, that both the {\tt diskbb} and {\tt bbody} components are Comptonized, is discussed in the next section. 

Before examining the double-seed scenario, we tested another single-seed model {\tt bbody+nthcomp[diskbb]}, which is identical to the modelings by, e.g., \citet{Church95},  assuming that only the disk photons are Comptonized. As shown in figure \ref{fig:fourspectra}(c), the fit was acceptable with a similar fit goodness of $\chi^2(\nu)=153.3(157)$, and yielded the best-fit parameters as $kT_{\rm bb}=2.16\pm0.06$ keV, $R_{\rm bb}=0.59\pm0.07$ km, $kT_{\rm in}=0.44\pm0.03$ keV, $R_{\rm in}=27\pm10$ km (assuming $\theta = 70^{\circ}$), $kT_{\rm e} = 9.1^{+4.7}_{1.6}$ keV, and $\tau = 4.7^{+1.2}_{-0.8}$. However, the fit implies a clear over-dominance of the (Comptonized) disk component. Actually, the fractional flux (uncorrected for inclination) carried by the blackbody component is $< 20\%$, in contradiction to the expectation that this component should carry more than half the total flux in high-inclination sources \citep{Mitsuda84}. Furthermore, the luminous Comptonized disk emission must be arising from a geometrically-thick and optically-thin accretion flow \citep[e.g.,][]{Shapiro76,Sakurai14}, which carries a dominant fraction of the overall mass accretion rate. Such a flow would inevitably Comptonize the NS blackbody photons, as well as the disk emission. This alternative modeling is thus considered unphysical, and is hence no longer considered in the present paper.

\subsection{Fit with a double-seed Comptonization model} 
\label{sec:model3}

The simplest form of the double-seed photon Comptonization is to assume that the same electron corona gives equivalent effects on both seed photons; this can be written in the form of {\tt nthcomp[diskbb+bbody]}. The values of $kT_{\rm e}$ and $\Gamma$ were thus constrained to be the same between the two Comptonization components. As shown in figure \ref{fig:fourspectra}(d) and table \ref{tab:totalspectra}, this model is also successful with $\chi^2(\nu)=151.8(157)$. Compared to the single-seed model employed in subsection \ref{sec:model2}, the NS blackbody component moves to a higher energy band, with $kT_{\rm bb}$ increasing and $R_{\rm bb}$ decreasing. The accretion disk steps back with a lower $kT_{\rm in}$ and a $\sim$ 60\% larger $R_{\rm in}$. Although the coronal parameters have become much less constrained, this double-seed modeling favors a corona of a higher $kT_{\rm e}$ and a lower $\tau$. Because cooler seed photons from the disk are now included, the average seed-photon temperature decreases, which requires more energy compensation from the corona, and hence a higher value of $kT_{\rm e}$. This increase in $kT_{\rm e}$ is compensated for by a decrease in $\tau$. 

In an attempt to disentangle the Comptonization between the disk component and the NS component, we also allowed the two Comptonization components to have separate values of $\tau$, but $kT_{\rm e}$ being the same. Then, the fit was improved only by $\Delta \chi^2=0.4$ for $\Delta \nu$ = -1, and $kT_{\rm e}=15.8(>1.53)$ keV was obtained. The NS component prefers a smaller spectral slope ($\Gamma_1=2.75^{+0.33}_{-0.19}$) than the disk component ($\Gamma_2=3.56^{+5.97}_{-1.13}$), and hence prefers a larger coronal optical depth but obviously with low significance. The upper limit on $\Gamma_2$ corresponds to an optical depth of $\sim$ 0.3, which indicates a very weak Comptonization effect on the disk emission. We further allowed the two Comptonization components to have separate $kT_{\rm e}$, but the fit was not improved.

\begin{table}
\begin{center}
\caption{Results of the model fittings to the 0.8--45 keV non-dip non-burst spectra of 4U 1915-05.$^{\ast}$}
\label{tab:totalspectra}
\begin{tabular}{ccccc}
\hline
\hline

Components  &  Parameters                     &  single-seed$^{\dag}$        &     double-seed$^{\ddag}$       \\
\hline

wabs        &  $N_{\rm H}(10^{22}$ cm$^{-2})$ &     $0.33\pm0.01$	     &     $0.36\pm0.01$	       \\ 
\hline
edge        &  $E_{\rm edge}$ (keV)           &     1.56(fixed)	             &     1.56(fixed)  	       \\
            &   optical depth                 &     $0.049\pm0.016$          &     $0.038\pm0.016$	       \\
\hline 
gabs        &  $E$ (keV)                      &     $6.68^{+0.08}_{-0.05}$   &     $6.68^{+0.11}_{-0.05}$      \\  
(Fe XXV K$\alpha$) &  $\sigma$ (eV)           &     $1.1^{+1.0}_{-0.7}$      &     $1.0 (>0.8)$    	       \\  
	    &   line depth ($10^{3}$)         &     $31 (>6.0)$	             &     $83 (>29.0)$   	       \\  
\hline
gabs        &  $E$ (keV)                      &     $6.95\pm1.39$	     &     $6.95^{+0.03}_{-0.05}$      \\  
(Fe XXVI K$\alpha$) &  $\sigma$ (eV)          &     $2.3 (>1.6)$ 	     &     $2.3 (>1.7)$		       \\  
	    &   line depth ($10^{3}$)         &     $100 (>5.4)$	     &     $892 (>4.2)$		       \\  
\hline
{\tt diskbb} &  $kT_{\rm in}$ (keV)           &     $0.70\pm0.01$	     &     $0.56\pm0.04$	       \\  
            &  $R_{\rm in}^{\S}$ (km)         &     $10.4\pm0.2$             &     $16.7\pm0.2$	               \\  
	    &  $F_{\rm disk}^{\|}$            &      2.13                    &      2.16                       \\
\hline
{\tt bbody} &  $kT_{\rm bb}$ (keV)            &     $1.31\pm0.03$	     &     $1.67\pm0.06$               \\  
            &  $R_{\rm bb}$ (km)              &     $2.19^{+0.03}_{-0.06}$   &     $1.25^{+0.03}_{-0.08}$      \\  
	    &  $F_{\rm bb}^{\|}$              &      1.83                    &      1.59                       \\
\hline
{\tt nthcomp} & $kT_{\rm e}$ (keV)            &     $9.4^{+3.3}_{-2.9}$      &     $40.1 (>14.0)$	       \\  
            &  $\Gamma$                       &     $2.53^{+0.10}_{-0.12}$   &     $2.99^{+0.17}_{-0.35}$      \\  
            &  $\tau$                         &     $3.6^{+1.2}_{-0.8}$	     &     $1.0 (<3.5)$	               \\ 
	    &  $F_{\rm c}^{\#}$               &       1.88                   &      2.31                       \\ 
\hline
Fit goodness  &  $\chi^2(\nu)$                &     155.5 (157)  	     &     151.8 (157)		       \\  
\hline 
\hline
\end{tabular}
\end{center}
$^{\ast}$ The 0.8-10 keV XIS spectrum and 12-45 keV HXD-PIN spectrum were fitted simultaneously. The quoted errors refer to 90\% confidence limits.\\
$^{\dag}$ The single-seed model corresponds to {\tt diskbb+nthcomp[bbody]}. \\
$^{\ddag}$ The double-seed model corresponds to {\tt nthcomp[diskbb+bbody]}. \\
$^{\S}$ $R_{\rm in}$ was corrected for the inclination factor $\sqrt{\cos(\theta)}$, in which $\theta=70^{\circ}$ was assumed. \\
$^{\|}$ $F_{\rm disk}$ and $F_{\rm bb}$ are the unabsorbed bolometric flux of {\tt diskbb} and {\tt bbody} components in 0.1-200 keV, regardless of the Comptonization effect. The unit is $10^{-10}$ erg cm$^{-2}$ s$^{-1}$. \\
$^{\#}$ $F_{\rm c}$ is the energy flux added to the blackbody components by the Comptonization in unit of $10^{-10}$ erg cm$^{-2}$ s$^{-1}$.  
  
\end{table}

\section{DISCUSSION and CONCLUSION}

\subsection{The Spectral State of 4U 1915-05}
\label{sec:discuss1}

Among the Suzaku archive of Galactic dipping LMXBs, 4U 1915-05 is the most luminous one. In order to identify the spectral state of this source, and further probe its accretion scheme, we analyzed its archived Suzaku data set acquired on 2006 November 8. The X-ray signal was detected up to $\sim$ 45 keV with a total exposure time of 39.1 ksec. Since the HXD-PIN vs. XIS hardness ratio was approximately constant (excluding the dips and a burst) as the source varied by a factor $\sim 2$, we analyzed the time-averaged non-dip spectra. The broad-band spectrum, shown in figure \ref{fig:fourspectra}, is more convex than those of non-dipping LMXBs in the hard state, and is more similar to those in the soft state. It has been fitted successfully by the canonical spectral model for LMXBs in the soft state \citep[{\tt diskbb+bbody},][]{Mitsuda84}, on condition that a strong Comptonization is applied either on the blackbody component (single-seed), or on the entire continuum (double-seed). The derived continuum parameters ($kT_{\rm in}$, $R_{\rm in}$, $kT_{\rm bb}$ and $R_{\rm bb}$) are all physically reasonable in either modeling. 

In terms of the {\tt diskbb+nthcomp[bbody]} model, let us compare the derived parameters of 4U 1915-05 with those of Aql X-1, a typical non-dipping LMXB, observed with Suzaku in the soft state \citep{Sakurai12}. The two objects both show $R_{\rm bb} \sim 1-2$ km, which implies that the blackbody emission arises from an equatorial belt-like region on the NS surface. They also commonly exhibit $R_{\rm in} \sim 10-20$ km, indicating that the inner edge of the accretion disk is close to the NS surface. These two parameters together give a constraint on the geometry of the corona as presented in figure \ref{fig:geometry}. The measured coronal $kT_{\rm e}$ of 4U 1915-05, $\sim$ 9 keV, is somewhat higher than that of Aql X-1 ($\sim$ 2-3 keV), but is significantly lower than those (several tens of keV) of LMXBs in the hard state \citep{Lin07,Sakurai14}.

Assuming the source distance of 9.3 kpc \citep{Yoshida93} and isotropic emission, an unabsorbed $0.8-45$ keV luminosity of 4U 1915-05 is $\sim 5.3\times10^{36}$ erg s$^{-1}$, which becomes $\sim 6\times10^{36}$ erg s$^{-1}$ via bolometric correction (here the inclination factor is not considered). The latter corresponds to $\sim$ 0.02 $L_{\rm edd}$, where $L_{\rm edd}=3.8\times10^{38}$ is the Eddington luminosity for a He-rich UCXB \citep{Kuulkers03}. The estimated Eddington ratio is reasonable, or somewhat lower for the soft state, because the soft-state vs. hard-state transition luminosity of LMXBs is generally considered to be $0.01-0.04$ $L_{\rm edd}$ \citep{Maccarone03,Matsuoka13}. However, the estimated intrinsic luminosity could be higher by a factor of up to 2, if the source inclination is higher than $\theta$ $\sim$ 70$^{\circ}$ \citep{Mitsuda84}.

\subsection{The Two Comptonization modelings}
\label{sec:discuss2}

As modifications of the simple Mitsuda model for 4U 1915-05 in the soft state, the single-seed and double-seed Comptonizing models, in the forms of {\tt diskbb+nthcomp[bbody]} and {\tt nthcomp[diskbb+bbody]} respectively, were both successful (figure \ref{fig:fourspectra} and table \ref{tab:totalspectra}). As the simpler of them, the single-seed model gave good constraints on the parameters of the Comptonizing corona, within typical 90\%-confidence errors of $\sim$30\%. The only demerit of this model is probably the value of $R_{\rm in}$ ($\sim 10~{\rm km}$), which is relatively small and may contradict to the single-seed assumption, that the disk photons are not scattered by the Comptonizing corona. Even if $R_{\rm in}$ is multiplied by a factor of $\xi \kappa^2$, where $\xi=0.412$ is a correction factor for the inner-boundary condition which was applied for black hole accretion \citep{Kubota98,Makishima00}, and $\kappa=1.7$ is a factor for the possible color hardening \citep{Shimura95}, $R_{\rm in}$ will increase only to $12.4\pm0.2$ km which is still relatively small. However, other factors may also affect $R_{\rm in}$: the source distance $D$ and inclination angle $\theta$, which are incorporated in the form of $R_{\rm in} \propto D/\sqrt{{\rm cos}(\theta)}$. If $D$ is $\sim$20\% larger and $\theta$ is $\sim$80$^{\circ}$ (section \ref{sec:introduction}), $R_{\rm in}$ increases to $\sim$ 20 km. Considering all these uncertainties, it is too premature to conclude that the single-seed modeling is unrealistic.

The double-seed Comptonization model is also successful, and gives physically reasonable parameters for both of the two seed-source components (table \ref{tab:totalspectra}). However, considering the results in section \ref{sec:model3}, a more realistic picture could be in between of the two modelings \citep[e.g.,][]{Sugizaki13}, that the disk blackbody emission is presumably Comptonized but the effect is weaker than that of the NS blackbody. Indeed, if we allowed the two components to have different $\tau$, the disk component prefers a smaller value (though statistically insignificant) than the other (section \ref{sec:model3}).

\subsection{The Inclination Effect on the Comptonization Strength}
\label{sec:discuss3}

As reviewed in section \ref{sec:introduction}, dipping LMXBs have generally been reported to have relatively harder spectra \citep{Church98,Iaria01,Oosterbroek01,Church99,Sidoli05}, in comparison with non-dipping LMXBs \citep{Gladstone07}. It is hence suggested that dippers, with higher inclination angles than normal ones, suffer stronger Comptonization effects, presumably because the Comptonizing corona extends along the disk plane rather than being spherical, as illustrated in figure \ref{fig:geometry}(b). Such an idea has already been discussed in black hole binaries, through a comparison of the high inclination source GRO J1655-40 with the low inclination source Cyg X-1 \citep{Makishima07}. Below, we carry out such an attempt, by quantitatively comparing 4U 1915-05 with non-dipping LMXBs in terms of their strength of Comptonization.

The strength of Comptonization is normally evaluated by the $y$-parameter in the form of $y=4kT_\textmd{\scriptsize{e}}\textmd{max}(\tau,\tau^2)/m_\textmd{\scriptsize{e}}c^2$, when $kT_{\rm e}$ is much larger than the seed photon temperature, $kT_{\rm s}$. However, for LMXBs in the soft state, the difference between $kT_{\rm e}$ and $kT_{\rm s}$ becomes rather small so that we can no longer assume $kT_{\rm e} \gg kT_{\rm s}$. Thus we may employ a new definition of $y$-parameter as $4(\tau+\tau^2/3)(kT_{\rm e}-kT_{\rm s})/m_{\rm e}c^2$, in which the term $(kT_{\rm e}-kT_{\rm s})/m_{\rm e}c^2$ represents the amount of energy transferred from the corona to the seed photons, and the scattering number is approximated as $\tau+\tau^2/3$ based in the {\tt nthcomp} code. The newly defined $y$-parameter of 4U 1915-05 is $y=0.50^{+0.05}_{-0.03}$ based on the single-seed model in which the Comptonization parameters are better constrained (section \ref{sec:model2}). This value appears at the upper limit of the recalculated y-parameters of normal soft-state LMXBs with similar $kT_{\rm bb}$ around 1.0-1.5 keV. Examples include Aql X-1 \citep[$y\sim0.22$,][]{Sakurai12}, 4U1608-522 \citep[$y\sim0.15-0.25$,][]{Tarana08}, and 4U 1735-44 \citep[$y\sim0.49$,][]{Muck13}. Thus 4U 1915-05, having a high inclination angle, is inferred to show stronger Comptonization effects than low and medium inclination LMXBs. 

Uniquely among dipping LMXBs, 4U 1915-05 is recognized to contain a helium rich donor star \citep{Nelemans06}. Since details of the Comptonization could depend on the chemical composition of the accreting matter, the above inference will become more secure, by comparing the Comptonization strength of this source with non-dipping LMXBs which also have helium donors. Among the $\sim15$ Galactic UCXBs, the non-dipping source 4U 1820-30 located at $6.4$ kpc \citep{Vacca86} is a good candidate, because it possesses a helium white dwarf donor star \citep{Stella87} and an inclination of $35^{\circ}$ - $50^{\circ}$ as estimated from strong orbital modulation of its ultraviolet flux \citep{Anderson97}. \citet{Bloser00} noticed that these two ultracompact LMXBs, i.e., 4U 1915-05 and 4U 1820-30, occupy similar spectral states, but in a color-color diagram 4U 1915-05 always exhibits larger hard color (the ratio of RXTE PCA count rates in the bands 9.7-16 keV and 6.4-9.7 keV). For quantitative investigation we have analyzed one Suzaku observation of 4U 1820-30 as detailed below.

\begin{figure*}
  \begin{center}
    \includegraphics[angle=0, width=0.6\textwidth]{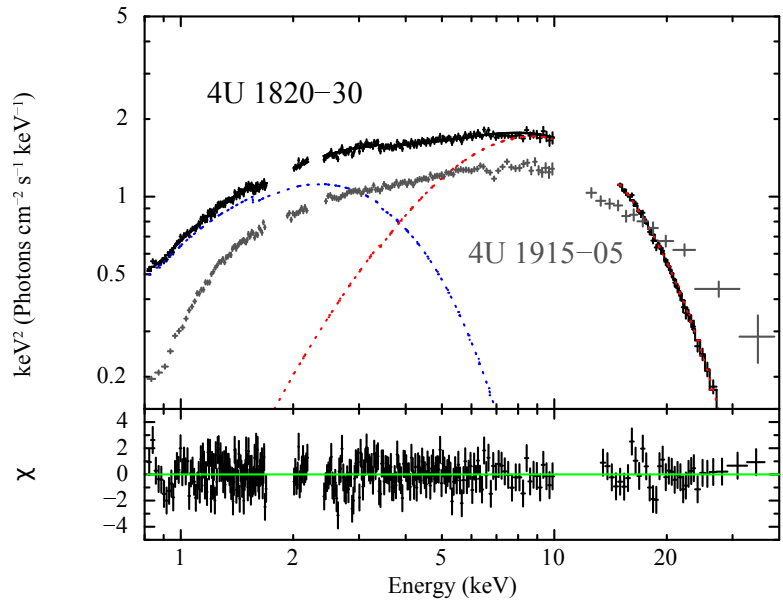} 
  \end{center}
  \caption{The $\upsilon F \upsilon$ plot of the Suzaku spectrum of 4U 1820-30, fitted with the same {\tt diskbb} (blue) plus {\tt nthcomp[bbody]} (red) model as used for 4U 1915-05. The spectrum of 4U 1915-05 (grey) is superposed with an arbitrary shift in normalization.}
\label{fig:4U1820}    
\end{figure*}

4U 1820-30 has been observed by Suzaku eight times from 2006 September 14 to 2009 October 20. The source was in the soft state in all the eight observations, and the last one (ObsID: 404069070) caught its dimmest and hardest phase among the eight, with a 0.8-10 keV XIS signal rate of $90.7\pm0.1$ cts s$^{-1}$, and a 15-40 keV PIN signal rate of $4.68\pm0.02$ cts s$^{-1}$. Hence we analyzed this data set in detail under similar procedure as described in section \ref{sec:observation}. The only difference is that we did not use XIS 3 because of its stronger pile-up effects in this observation. The XIS source region was chosen to be an annulus between $0^{\prime}$.5 and $3^{\prime}$.5, and the background region was an outer annulus between $3^{\prime}$.5 and $5^{\prime}$. The cleaned total exposure time was 8.7 ksec for XIS (CLK mode = `burst') and 16 ksec for HXD-PIN. Since no dip or burst event was detected in this observation, the XIS and PIN spectra, shown in figure \ref{fig:4U1820}, were accumulated over the total cleaned exposure time. 

In figure \ref{fig:4U1820}, we superpose the non-dip spectrum of 4U 1915-05 (the same as figure \ref{fig:fourspectra}) with an arbitrary shift in normalization. The two spectra are similar below 10 keV, but 4U 1915-05 exhibits a much harder PIN spectrum. We fitted the 0.8-40 keV spectrum of 4U 1820-30 by the same {\tt diskbb+nthcomp[bbody]} model and list the obtained fitting parameters in table \ref{tab:4U1820para}. The fitting has been successful with $\chi^2(\nu)=296.0(277)$ and the obtained model parameters are all reasonable for a soft-state LMXB. The derived Comptonization $y$-parameter of 4U 1820-30 is $y=0.27^{+0.16}_{-0.15}$, which is smaller than the $y$-parameter of 4U 1915-05 ($y=0.50^{+0.05}_{-0.03}$) at the 90\% confidence level. Moreover, in table \ref{tab:totalspectra} and \ref{tab:4U1820para}, we have calculated the model dependent bolometric fluxes of the optical-thick components ($F_{\rm bb}$ and $F_{\rm disk}$), and the flux added to them by Comptonization (the coronal flux $F_{\rm c}$). The fraction of the coronal flux contributing to the total flux is 50\% in 4U 1915-05 (based on the single-seed model), while only 17\% in 4U 1820-30. Thus, the spectrum of 4U 1916-05 is more strongly Comptonized than that of 4U 1820-30. This is likely to be caused by their different inclinations, because both have helium-rich donors. These results suggest that the corona is oblate, at least in these two objects.  

In conclusion, we have shown that 4U 1915-05 was in the soft state during the Suzaku observation, and its 0.8-45 keV spectrum can be described by the standard spetral model for LMXBs in the soft state, except that the Comptonization is significantly stronger. 

We thank the supports from the Japan Society for the Promotion of Science (JSPS) to foreign post-doctoral researchers. The awarded Grant-In-Aid number is 24-02321.

\begin{table}
\begin{center}
\caption{Results of the {\tt diskbb+nthcomp[bbody]} fitting to the Suzaku XIS and HXD-PIN spectra of 4U 1820-30.$^{\ast}$}
\label{tab:4U1820para}
\begin{tabular}{ccc}
\hline
\hline

Component   &  Parameter                      &     Value                      \\
\hline

wabs        &  $N_{\rm H}(10^{22}$ cm$^{-2})$ &     $0.06\pm0.01$	       \\ 
\hline
edge        &  $E_{\rm edge}$ (keV)           &      1.56(fixed)	       \\
            &   optical depth                 &     $0.066\pm0.012$            \\
\hline
edge        &  $E_{\rm edge}^{\dag}$ (keV)    &     $3.4\pm0.1$	               \\
            &   optical depth                 &     $0.027\pm0.013$            \\
\hline
{\tt diskbb}&  $kT_{\rm in}$ (keV)            &     $0.97\pm0.02$              \\  
            &  $R_{\rm in}^{\ddag}$ (km)      &     $10.8\pm0.4$               \\  
	    &  $F_{\rm disk}^{\S}$            &      3.85                      \\
\hline
{\tt bbody} &  $kT_{\rm bb}$ (keV)            &     $1.87\pm0.18$              \\  
            &  $R_{\rm bb}$ (km)              &     $3.2\pm0.2$                \\  
	    &  $F_{\rm bb}^{\S}$              &      3.09                      \\
\hline
{\tt nthcomp} & $kT_{\rm e}$ (keV)            &     $3.7^{+1.8}_{-0.4}$        \\  
            &  $\Gamma$                       &     $2.65^{+1.28}_{-0.46}$     \\  
            &  $\tau$                         &     $6.2^{+2.4}_{-3.4}$        \\ 
	    &  $F_{\rm c}^{\S}$               &      1.39                      \\ 
\hline
Fit goodness  &  $\chi^2(\nu)$                &     325.7 (274)  	       \\  
\hline 
\hline
\end{tabular}
\end{center}
$^{\ast}$ The 0.8-10 keV XIS spectrum and 15-40 keV HXD-PIN spectrum were fitted simultaneously. The quoted errors refer to 90\% confidence limits.\\
$^{\dag}$ Another edge was found around 3.4 keV, which is probably an ionized Argon K-edge.\\
$^{\ddag}$ $R_{\rm in}$ was corrected for the inclination factor $\sqrt{\cos(\theta)}$, in which $\theta=45^{\circ}$ was assumed. \\
$^{\S}$ Definition of $F_{\rm disk}$, $F_{\rm bb}$ and $F_{\rm c}$ are the same as in table \ref{tab:totalspectra}, however the unit is $10^{-9}$ erg cm$^{-2}$ s$^{-1}$.   
\end{table}

\end{document}